# Survival of the cheapest: How proteome cost minimization drives evolution

Kasper P. Kepp*

*Technical University of Denmark, DTU Chemistry, Kemitorvet 206,*

*DK-2800 Kongens Lyngby, Denmark.*

* Corresponding e-mail: kpj@kemi.dtu.dk

**Running title: Survival of the cheapest**

**Word count: 10,169 words**

**Abstract**

Darwin's theory of evolution emphasized that positive selection of functional proficiency provides the fitness that ultimately determines the structure of life, a view that has dominated biochemical thinking of enzymes as perfectly optimized for their specific functions. The 20$^{th}$-century modern synthesis, structural biology, and the central dogma explained the machinery of evolution, and nearly neutral theory explained how selection competes with random fixation dynamics that produce molecular clocks essential e.g. for dating evolutionary histories. However, the quantitative proteomics revealed that fitness effects not related to functional proficiency play much larger roles on long evolutionary time scales than previously thought, with particular evidence that some universal biophysical selection pressures act via protein expression levels. This paper first summarizes recent progress in the 21$^{st}$ century towards recovering this universal selection pressure. Then, the paper argues that proteome cost minimization is the dominant, underlying "non-function" selection pressure controlling most of the evolution of already functionally adapted living systems. A theory of proteome cost minimization is described and argued to have consequences for understanding evolutionary trade-offs, aging, cancer, and neurodegenerative protein-misfolding diseases.

**Table of contents**

**Introduction**

Protein evolution occurs via mutations that change the composition or expression of the proteome of a population, sometimes by random nearly-neutral drift, and sometimes via selection pressures imposed by the habitat.(Bajaj & Blundell 1984; DePristo *et al.* 2005; Goldstein 2008; Hurst 2009; Pál *et al.* 2006; Worth *et al.* 2009) After Darwin's theory of natural selection, Mendel's laws of inheritance, the modern synthesis of the 20$^{th}$ century, and the rise of structural biology and the central dogma, we know that nature selects favorable traits if their impact outweighs the random fixation dynamics, and we know how these changes are actualized via mutations in the DNA that translate to the proteome. Remaining major questions are: 1) How important is selection vs. random drift and can we predict their relative importance?(Blundell & Wood 1975; Hurst 2009; Kimura 1962; Ohta 1992) 2) What are the molecular properties selected for, and are they universal?(Hurst 2009; Liberles *et al.* 2012; Lobkovsky *et al.* 2010) 3) How do we describe accurately and completely the evolution of populations from the arising mutation in the gene, via the molecular property of the protein, to its fixation and ultimate effect on the population? According to this view, the ultimate goal of biology is to bridge the genome, proteome, phenotype, and population together in one quantitative and predictive theory that explains the history, present, and future of biological structure on this planet.

Darwin's theory of evolution emphasized that positive selection of functional proficiency provides the fitness that ultimately determines the structure of life (survival of the fittest). In the sixties, the observation of nearly constant evolution of homologous proteins(Margoliash 1963; Zuckerkandl & Pauling 1965, 1962) led to the theory of (nearly) neutral evolution showing that most fitness effects are too subtle to dominate the random fixation dynamics of the population, thus producing an almost constant rate of evolution.(Kimura 1962; Ohta 1992) This widely presumed molecular clock is essential for dating phylogenies and evolutionary histories.(Kumar & Subramanian 2002; Meredith *et al.* 2011; Yi *et al.* 2002; Zuckerkandl & Pauling 1965) When applied to single individuals, variations in the clock specific to the mutated site are widely assumed to indicate pathogenicity of a human gene variant.(Flanagan *et al.* 2010; Ng & Henikoff 2003; Shihab *et al.* 2013; Tang *et al.* 2019a)

This review concerns the question: What drives protein evolution on most time scales where the function is already nearly optimal after positive selection? To address this question we must first discuss the typical properties of proteins. Proteins vary by three orders of magnitude in length (from tens to ten thousands of amino acids), they vary structurally via thousands of folds,(Bajaj & Blundell 1984; Koonin *et al.* 2002; Mirny & Shakhnovich 1999; Qian *et al.* 2001) and by perhaps 5−7 orders of magnitude in abundance in eukaryotic cells.(Beck *et al.* 2011; Jansen & Gerstein 2000; Milo 2013) In stark contrast to these enormous variations, proteins across all domains of life are marginally stable in a narrow range of perhaps 30−100 kJ/mol, barely preventing denaturation.(DePristo *et al.* 2005; Goldstein 2011) There are three possible origins of this phenomenon: Marginal stability is a selected beneficial trait, it arises form random mutation-selection dynamics, or it reflects stability-constrained functional optimization. In the first case, marginal stability ensures efficient turnover of aged and damaged proteins and reuse of amino acids; a too stable fold may be hard to degrade by proteases of the proteasome or even the highly acidic lysosome. In the second case, since mutations arise randomly and anything random done to an optimized system tends to reduce optimality, protein stability is constantly challenged by mutations that destabilize by perhaps 5 kJ/mol on average,(Tokuriki *et al.* 2007) and responsive selection keeps the protein stable.(Goldstein 2011; Taverna & Goldstein 2002) If so, marginal stability is not a selected trait but a consequence of the predominance of random drift, with mutation-selection dynamics constantly playing out near the denaturation threshold. Third, optimization of function occurs under the constraint of preventing denaturation. If so, marginal stability is not a selected trait or a consequence of random drift but reflects maximal trading of stability for function by investing protein fold free energy to minimize transition state barriers of enzymes.(Warshel 1998) Each explanation does not exclude the others, as trade-offs and drift depend greatly on the protein, phenotype, and population as discussed below.

The relative rate of incorporating non-synonymous relative to synonymous substitutions (dN/dS) is a key indicator of the speed of evolution that varies by many orders of magnitude between sites and proteins.(Drummond *et al.* 2005; Gillespie 1984, 1986; Zuckerkandl & Pauling 1965) In the limit where the normalized dN/dS ratio approaches 1, amino acid changes occur as fast as codon substitutions that preserve the type of amino acid, i.e. no net selection pressure acts on the mutation, and its fixation is subject to the random chance of the population and mating dynamics (*neutral evolution*).(Fay *et al.* 2002; Kimura 1991) If the variations in evolutionary rates are small, the evolution is nearly neutral(Ohta 1992). Correspondingly, large values of dN/dS indicate selection pressure towards a new fitness optimum (*adaption or positive selection*).(Hurst 2009)

The major driver of protein evolution was previously thought to be the functional proficiency of the protein, e.g. the turnover, $k_{cat}/K_M$, for the natural substrate of an enzyme.(Hurst 2009; Soskine & Tawfik 2010) The connectivity of many proteins (i.e. the extent of their involvement in diverse biochemical pathways) has been suggested to slow their rate of evolution, consistent with functional constraints on evolution.(Fraser *et al.* 2002; Hahn & Kern 2004; Wall *et al.* 2005) However, data set choices affect these conclusions, and in many cases selection pressures not relating to the protein's function may counter this tendency.(Bloom & Adami 2003, 2004; Drummond *et al.* 2005) It is equally controversial whether more dispensable (non-essential) proteins evolve faster(Hirsh & Fraser 2001; Jordan *et al.* 2002) or not(Bloom & Adami 2003; Hurst & Smith 1999). It is now recognized that, although protein function is of course still essential, it plays a much smaller role in evolution, in particularly on long timescales, than previously thought.(Bloom & Adami 2003, 2004; Drummond *et al.* 2005; Hurst & Smith 1999; Lobkovsky *et al.* 2010; Wylie & Shakhnovich 2011) During early evolution, fierce competition produced major selection pressures and evolutionary innovations in prokaryotes, and later in the early evolution of eukaryotes;(Lane & Martin 2010; Sousa *et al.* 2013) the rise of the eukaryote heralded major biochemical innovations largely driven by advantages relating to size and metabolism.(Lane 2011) Under these conditions, fitness probably reflected the ability to harvest energy and chemical components and use these to produce offspring.(Lane & Martin 2010; Sousa *et al.* 2013)

The subsequent long periods of relatively stable evolution have seen active sites of proteins highly conserved by purifying selection near their fitness optima(Blundell & Wood 1975; Casari *et al.* 1995) and most of the interesting variation in protein evolution rates comes from other sites typically subject to nearly neutral fixation dynamics. These nearly-neutral substitutions probably dominate most recent evolution(Ohta 1992), except unusual speciation and diversification events due to spontaneously arising new functions or selection pressures. For the same reason, almost all protein evolution involves sequence variations that maintain the already adopted, highly conserved fold structure.(Worth *et al.* 2009) The nearly neutral sites may contribute to optimal translational efficiency under favorable growth conditions.(Andersson & Kurland 1990; Ikemura 1985) Selection at the gene level for translational efficiency and precision(Akashi 2003; Andersson & Kurland 1990; Drummond *et al.* 2005; Marais & Duret 2001) is evident e.g. from codon bias and t-RNA isoforms.(Kanaya *et al.* 1999; Robinson *et al.* 1984; Tuller *et al.* 2010)

**The main determinants of evolution rate**

To understand the main drivers of evolution we must first understand the protein properties that mostly determine evolutionary rates more broadly in proteins and on longer time scales. This rate is also used to classify and predict the functional impact of human variants e.g. in relation to disease.(Capra & Singh 2007; Glaser *et al.* 2003; Tang *et al.* 2019b; Thusberg *et al.* 2011) **Table 1** gives an overview of the most important relationships between a protein's properties and its evolution rate. As easily verified from sequence alignment, active sites in proteins are highly conserved due to strong purifying selection, since random deleterious mutations impair fitness more in highly optimized parts of the protein. Related to this, solvent-exposed sites in contrast evolve faster than average, consistent with their typically small functional and structural effects on the overall protein.(Goldman *et al.* 1998; Overington *et al.* 1992; Ramsey *et al.* 2011)

The strongest descriptor of evolutionary rate is protein abundance or equally, mRNA levels, as these correlate;(Gygi *et al.* 1999) it typically spans 5−7 orders of magnitude in eukaryotes.(Beck *et al.* 2011; Ghaemmaghami *et al.* 2003; Jansen & Gerstein 2000; Milo 2013) High expression is associated with slower protein evolution in both prokaryotes(Rocha & Danchin 2004; Sharp 1991) and eukaryotes(Pál *et al.* 2001), including mammals(Jordan *et al.* 2004; Zhang & Li 2004), a phenomenon known as the E-R anti-correlation.(Bloom *et al.* 2006a; Drummond *et al.* 2005) Protein expression may explain half of the evolutionary rate variation in yeast(Drummond *et al.* 2006) indicating a universal driving force of evolution. This remarkable relationship has been studied with many biophysical models focusing on protein stability, misfolding avoidance, packing, and flexibility.(Dasmeh *et al.* 2014b; Geiler-Samerotte *et al.* 2011; Kepp & Dasmeh 2014; Liberles *et al.* 2012; Lobkovsky *et al.* 2010; Serohijos *et al.* 2012; Sikosek & Chan 2014; Wylie & Shakhnovich 2011; Yang *et al.* 2012) All-else-being-equal, a protein's fitness impact on the organism should be proportional to its cellular abundance regardless of the specific selection pressure. Thus, any fitness function that scales with protein abundance may seem reasonable. Such models can explain about 60% of site-variations in evolutionary rate.(Echave *et al.* 2016; McInerney 2006) A protein's thermodynamic stability (free energy of folding) is thought to be an important contributing determinant of evolutionary rate, although its relation to fitness so far has entered via its relation to the copy number of misfolded proteins, assuming one-step unfolding.(Dasmeh *et al.* 2014a; Serohijos *et al.* 2012) These ideas are expanded further below. To summarize the tendencies of **Table 1**, compared to the average protein and all-else-being-equal, the slowly evolving protein tends to be highly expressed, intracellular, smaller than average, and have a higher functional density, i.e. more important sites *relatively to its size*.

The E−R anti-correlation has been explained(Drummond & Wilke 2008, 2009) as a selection against inefficient translation leading to misfolded proteins, as these are widely assumed to be toxic. Protein synthesis is inherently error-prone, and translation operates with typical missense error rates of 1/1000 to 1/10,000.(Drummond & Wilke 2008) Considering the typical lengths (~100−1000) and total abundance of proteins ($10^8$) in eukaryotic cells, one can expect $10^{10}$−$10^{11}$ protein-incorporated amino acids to exist at any time. Without error correction this could imply the constant existence of $10^6$−$10^8$ erroneous amino acids in a typical eukaryote cell. This would make translation-error induced proteome variation of similar importance as typical, mostly heterozygote, natural sequence variation in a population. This of course raises the question how much of the actual observed proteome variation is due to genetic inheritance, somatic mutations, and translation errors. To be sure, one needs to sequence each gene and protein many times for several cells. Regardless of this complication, it is clear that the proteome varies much more in composition than implied by genetic variance alone.

Considering this, since the typical non-native residue destabilizes by ~5 kJ/mol,(Tokuriki *et al.* 2007) it is possible that as much as 10% of a proteome could be less stable than commonly assumed purely from wild type sequence. For a cell with $10^8$ proteins, this implies that $10^7$ protein copies are randomly destabilized and subject to higher turnover that expected from their wild type sequence. Post-translational modifications and specific degrons further diversify the proteome and complicate turnover further. Considering this, the additional destabilization from new arising mutations will aggravate costs only if the affected protein is quite abundant or subject to high turnover.

According to Drummond and Wilke(Drummond & Wilke 2008), selection acts to reduce the toxicity or functional loss due to misfolding. According to the PCM theory, as explained below, selection acts to reduce the expensive ATP cost of turnover of these erroneous protein copies, mostly because of the synthetic cost of replacing the erroneous copy with a new functional and stable wild type copy. This distinction is very important, since in the latter case the consequences of protein misfolding directly relate to the well-established quantitative facts of bioenergetics. If the misfolded protein is selected against, regardless of the reason, highly expressed proteins are under stronger selection pressure since the copy number of misfolded proteins $U_i$ scales with the total abundance of the protein $A_i$. Drummond and co-workers suggested a fitness function $\Phi$ depending exponentially on the total copy number of all misfolded proteins $U = \sum U_i$, with an unknown scaling constant $c$:(Drummond & Wilke 2008)

$$\Phi \propto \exp(-cU) \tag{1}$$

The constant $c$ can be derived from fundamental and simple assumptions and related directly to the cost of protein turnover, as discussed below.

**The theory of proteome cost minimization**

Darwin's theory of selection and the theory of nearly neutral evolution(Kimura 1962, 1991; Ohta 1992) together explain evolution as a process of selection and drift, whereas structural biology explains the molecular language of evolution via the central dogma. However, a complete theory of evolution requires us to also know the property of the evolving protein that contributes to the organism phenotype, why it contributes, to what extent it contributes, and how this affects the wider evolution of the population in its ecological and historical context. As discussed extensively in the literature, it is increasingly clear that the functional traits selected for in classical positive Darwinian evolution have relatively little importance in many cases relative to other, partly hidden and perhaps universal properties of the proteins.(Bloom & Adami 2003, 2004; Drummond *et al.* 2005; Hurst & Smith 1999; Lobkovsky *et al.* 2010; Wylie & Shakhnovich 2011)

The most obvious universal property subject to selection pressure is arguably the cellular energy state. Before the era of structural biology and proteomics, Boltzmann(Boltzmann 1886) and Schrödinger(Schrödinger 1944) already speculated that life characteristically represents a well-defined organized (low-entropy) structure that maintains a thermodynamic non-equilibrium state relative to its high-entropy surroundings by constant energy turnover and associated heat dispersion. By this definition, expansion of life (fitness) implies expansion of this energy turnover. Lotka applied these ideas to Darwin's selection theory via his maximum power principle, arguing that evolution occurs by selection of the most energy-efficient organisms.(Lotka 1922) These ideas were then expanded into a much broader ecological view by Odum.(Odum 1988) Thermodynamically, the system most capable of maintaining its structure by energy dissipation and with the ability to grow and reproduce these structures will prevail over other similar systems, and thus, be most fit.

The theory presented below was inspired by these views and further supported by the observations of consistent cost-bias in amino acid use across all kingdoms of life first discovered by Akashi and Gojobori.(Akashi & Gojobori 2002) These findings were confirmed by Swire(Swire 2007) and later explained in a well-argued fitness model by Wagner who showed, among other things, that gene duplications are highly selected against in terms of cellular energy costs.(Wagner 2005) The theory builds substantially on Wagner's seminal quantitative considerations(Wagner 2005) and the important considerations of Brown, Marquet, and Tape(Brown *et al.* 1993) who used Lotka's ansatz to explain mass and size optima of biological taxa in terms of evolutionary fitness caused by the different scaling of metabolic rates and reproductive rates with mass. The theory's central ansatz, inspired by these minds, is as follows: "*Fitness is proportional to the energy per time unit available for reproduction after subtracting (proteome) maintenance costs*". Since fitness always has to be measured relative to a wild type after an instant of time, the energy of interest becomes a power (measured in watt. or J s$^{-1}$) as in Lotka's original thinking, and as such directly relates to the respiration rate of the organism, as discussed below.

The mechanistic basis for the theory is that i) protein degradation increases many-fold with the lack of structure and partial unfolding in protein copies,(Gsponer *et al.* 2008) and ii) the cost of protein turnover is more than half of total metabolic costs in growing microorganisms,(Harold 1987) and at least 20% in humans.(Waterlow 1995) Accordingly, any increase in these costs reduces the energy available for other energy-demanding processes, notably reproduction (fitness) of microorganisms(Dasmeh & Kepp 2017) and cell signaling (cognition)(Kepp 2019) in higher organisms. One of many implications of the theory is that selection against misfolded proteins and toxicity of misfolding proteins measured in cell viability assays *is not due to a specific toxic molecular mode of action as widely assumed, but to the generic ATP burden of turning over the misfolded proteins within the cell*.(Kepp 2019)

In its simplest form, which is easily expanded, we assume a life cycle of a protein *i* as:

$$mRNA_i \xrightarrow{k_{s_i}} F_i \underset{k_{2_i}}{\overset{k_{1_i}}{\rightleftharpoons}} U_i \xrightarrow{k_{d_i}} D_i \tag{2}$$

$F_i$ represents the copy number of folded proteins, $U_i$ represents misfolded proteins, and $D_i$ represent the degradation products, many of which are recycled for use in other proteins; the rate constant of each process is specific to the protein in question. Since the ultimate selection pressure acts only on $U_i$, one can easily relax the assumption of one-step unfolding to account for complex situations.

$k_{d_i}$ is the rate constant (in units of protein molecules per second) for degrading the misfolded protein copies. The *in vivo* rate constants reflect the half-life (t$_{½}$) of the fully folded protein, and can thus be written at steady state as:

$$k'_{d_i} = \frac{k_{d_i} k_{1_i}}{k_{2_i}} = \frac{k_{d_i}}{K_{f_i}} = \frac{\ln 2}{t_{½}} \tag{3}$$

which varies substantially with the protein $i$, giving half lives from minutes to days.(Hargrove & Schmidt 1989) Accordingly, $k'_{d_i}$ is typically of the order of $10^{-4}$ s$^{-1}$ but with order-of-magnitude variations, whereas $k_{d_i}$ acts directly on already misfolded protein and represents the rate of protein degradation if the chemical activation barrier to unfolding has been removed. Thus, $k_{d_i}$ is limited by the number of active proteases, the diffusion and proper orientation of the exposed peptide bond, and the actual $k_{cat}/K_M$ of the proteases, with an upper limit of perhaps $10^6$ to $10^8$ M$^{-1}$s$^{-1}$ per peptide bond hydrolysis.(Bar-Even *et al.* 2011; Wolfenden & Snider 2001) In terms of steady-state turnover, misfolded proteins are immediately targeted for degradation(Gsponer *et al.* 2008) and recruited by the ubiquitin-proteasome pathway that takes the protein out of the pool, and thus this process is not rate-limiting the overall protein flux but arguably operates near the diffusion limit.

Assuming one-step misfolding, $U_i$ is related to the folding free energy of the protein $\Delta G_i = -RT \ln(K_{f_i})$ via the equilibrium constant $K_{f_i} = F_i / U_i$:

$$U_i = A_i \left( \frac{1}{1+\exp\left(\frac{-\Delta G_i}{RT}\right)} \right) \approx A_i \exp\left(\frac{\Delta G_i}{RT}\right) \tag{4}$$

The last expression follows if there are many more folded than unfolded copies of the protein, which is almost always the case. Since folding equilibrium constants easily reach $10^{11}$ for a protein of typical stability (65 kJ/mol at 37°C), the number of misfolded proteins at any given time is typically negligible, as they are immediately subject to turnover. Reasonable experimental values of $k_{d_i} = 10^7$ s$^{-1}$, $K_{f_i} = 10^{11}$, and $k'_{d_i} = 10^{-4}$ s$^{-1}$ satisfy the relationship in Equation (3) and thus justify the use of Equation (2).

Equation (4) is well established and was first used in a fitness function by Bloom et al.(Bloom *et al.* 2004) and has been specifically used to explain some of the E-R anticorrelation(Serohijos *et al.* 2012) and additional variations in evolutionary rates.(Dasmeh *et al.* 2014a) The advantage of this expression is that we can relate the number of misfolded protein copies, which is the property selected upon, directly to the total copy number $A_i$ of the protein within the cell and to its thermodynamic stability, via the free energy of folding $\Delta G_i$ (a negative number in kJ/mol). $RT$ is the thermal energy of the cell, and thus temperature enters directly as a fundamental physical parameter determining proteome $U_i$ and ultimately cellular proteome costs and fitness, as discussed further below.

The critical step is now to write the fraction of the total respiration rate (in watt, or J s$^{-1}$) of the cell due to the maintenance of a single protein $i$:

$$dE_{m,i}/dt = A_i \exp\left(\frac{\Delta G_i}{RT}\right) k_{d_i} N_{aa_i}(C_{s_i} + C_{d_i}) \tag{5}$$

In this equation, in addition to the parameters already described above, $N_{aa_i}$ represents the number of amino acids in the protein $i$, and the cost constants $C_{s_i}$ and $C_{d_i}$ describe the average synthetic and degradation cost per amino acid in protein $i$ in units of J.(Kepp & Dasmeh 2014)

For the whole proteome of the cell, we can write the *total* cost per time unit as the sum of the costs of maintaining steady-state folded protein copy numbers within the cell:

$$dE_m/dt = \alpha \sum_i dE_{m,i}/dt = \alpha \sum_i A_i \exp\left(\frac{\Delta G_i}{RT}\right) k_{d_i} N_{aa_i}(C_{s_i} + C_{d_i}) \tag{6}$$

Importantly, we see that the total energy costs scale with $A_i$. Since $A_i$ varies substantially for different proteins, e.g. from zero to a million, some proteins are much more important to the cell's energy budget than others. The scaling constant $\alpha$ represents the activity of the proteasome, which may be controlled with proteasome inhibitors, but a slight expansion of this expression can be done to ($\alpha + \beta + …$) taking into account the contributions of various degradation pathways (lysosome, proteasome, effects of N-end rule, etc.) to the overall turnover. **Figure 1** summarizes some typical values for the parameters of the model applicable to eukaryote cells.

**Selection dynamics of PCM**

To understand how protein turnover costs affect evolution, we now use the central ansatz that fitness scales with the energy available for reproduction $dE_r/dt$ after subtracting the proteome costs of Equation (6) from the total energy available to the cell either by production or supply, $dE_t/dt$, divided by the respiration rate needed to run an individual, also taken to $dE_t/dt$:

$$\Phi = \frac{dE_r/dt}{dE_t/dt} = \frac{dE_t/dt - dE_m/dt}{dE_t/dt} = 1 - \frac{dE_m/dt}{dE_t/dt} \tag{7}$$

The division by $dE_t/dt$ is not strictly required as it cancels out in selection coefficients, but formally ensures a dimensionless fitness function. For simplicity, we ignore the non-proteome energy costs since the purpose is to show that the cost of the proteome exerts a major effect on evolution by itself. Assuming that the total energy production is constant for all competing cells, minimization of $dE_m/dt$ maximizes fitness. When a new mutation arises in protein $i$, the selection coefficient is:

$$s_i(M) = \frac{\Phi_i(M)}{\Phi_i(WT)} - 1 = \frac{\Phi_i(M) - \Phi_i(WT)}{\Phi_i(WT)} = \frac{dE_m/dt(WT) - dE_m/dt(M)}{dE_t/dt(WT) - dE_m/dt(WT)} \tag{8}$$

For clarity, we have assumed that the mutation only affects maintenance turnover costs and not energy production, and thus the total energy produced is the same before and after mutation and cancels in Equation (8). If we further neglect epistasis, selection only acts on the mutated protein $i$:

$$s_i(M) = \frac{A_i \exp\left(\frac{\Delta G_i}{RT}\right) k_{d_i} N_{aa_i}(C_{s_i} + C_{d_i})(WT) - A_i \exp\left(\frac{\Delta G_i}{RT}\right) k_{d_i} N_{aa_i}(C_{s_i} + C_{d_i})(M)}{dE_t/dt(WT) - A_i \exp\left(\frac{\Delta G_i}{RT}\right) k_{d_i} N_{aa_i}(C_{s_i} + C_{d_i})(WT)} \tag{9}$$

This selection coefficient is a function only of protein properties, scaled by the general energy spent for reproduction of the organism, $dE_r/dt(WT)$, which can be taken as a constant of the order of $10^{-11}$ J s$^{-1}$.(Harold 1987) It is perhaps more convenient to write Equation (9) in terms of copy numbers and half lives (t½) which can be measured in live cells:

$$s_i(M) = \frac{A_i N_{aa_i}(C_{s_i} + C_{d_i}) \ln 2/t_{½}(WT) - A_i N_{aa_i}(C_{s_i} + C_{d_i}) \ln 2/t_{½}(M)}{dE_r/dt(WT)} \tag{10}$$

where we have used the relationship:

$$k'_{d_i} = \frac{\ln 2}{t_{½}} = \exp\left(\frac{\Delta G_i}{RT}\right) k_{d_i} \tag{11}$$

For a haploid organism, the probability of its fixation $P_{fix}$ is approximately:(Kimura 1962; Ohta 1992)

$$P_{fix} = \frac{1 - e^{-s_i}}{1 - e^{-s_i N}} \cong \frac{s_i}{1 - e^{-s_i N}} \tag{12}$$

where $N$ is the effective population size, and the last term comes from expanding the exponential of the small $s_i$. For neutral evolution, as $s_i \rightarrow 0$, $P_{\text{fix}} \rightarrow 1/N$, and does not depend on any properties of the protein. At significant positive selection, $s_i N$ is large, $s_i$ is positive, and $P_{\text{fix}} \rightarrow s_i$. Very similar behavior applies to diploid organisms with slightly different factors of 2 and 4.(Kimura 1962)

The simple kinetic scheme assumed for the PCM model is highlighted in **Figure 2A**. From Equation (10), considering the variations in the parameters, most of the proteome cost selection occurs by affecting the ratio $A_i / t_{½}$. Mutations that reduce the half life of abundant proteins are thus particularly selected against. The typical behavior of $P_{\text{fix}}$ with $N$ and $s_i$ is shown in **Figure 2B**. The absolute rate of evolution $\omega$ scales with the mutation rate and the probability of fixating new arising mutations:

$$\omega = uNP_{fix} = u\frac{s_i}{1-e^{-s_i N}} \tag{13}$$

where $u$ is the absolute mutation rate (the rate of nonsynonymous amino acid substitutions in real time in a protein copy of the total population); this expression can be expanded by life history variables such as generation time,(Martin & Palumbi 1993) but this is beyond the scope here, as the proportionality of Equation (13) generally applies, and $P_{\text{fix}}$ thus measures evolution rate. For an optimized evolutionary system, a typical arising mutation has a negative selection coefficient; if small relative to $1/N$, it is subject to random fixation drift. From Equation (13), such mutations will reduce the probability of fixation (and evolution rate) in proportion to the size of the negative selection coefficient. **Figure 2B** also illustrates why the molecular clock is generally successful at dating phylogenies, because 90% of randomly occurring mutations in the relevant selection-fixation space of **Figure 2B** are subject to neutral evolution.

In order to understand the slow evolution of abundant proteins discussed in the literature,(Bloom *et al.* 2006a; Drummond *et al.* 2005; Drummond & Wilke 2008) we should identify low values of $P_{\text{fix}}$ in the evolution rate space of **Figure 2B**. Most arising mutations (**Figure 2B**) remain subject to nearly neutral evolution. However, more extreme selection coefficients will occur for highly abundant proteins, since the selection coefficient of a new arising mutation in a protein scales with the abundance and turnover rate of the affected protein. In contrast, less abundant proteins will typically have numerically smaller selection coefficients at any given effective population size. The next section gives a quantitative estimate of the fixation probabilities.

**Typical PCM selection pressures and fixation probabilities for yeast**

**Table 2** summarizes some typical selection scenarios in yeast cells. A typical yeast cell respires at ~1 J s$^{-1}$ g$^{-1}$ and has a mass of 3·10$^{-11}$ g, giving $dE_t/dt \approx 3 \cdot 10^{-11}$ J s$^{-1}$. $C_{d_i}$ is perhaps 1 ATP per peptide bond or 30 kJ/mol.(Benaroudj *et al.* 2003) The biosynthetic costs of the amino acids vary from 10−80 ATP;(Wagner 2005) the average amino acid composition of the yeast proteome gives ~25 ATP, or 750 kJ/mol as typical. If half of the amino acids are recycled, neglecting amino acid transport cost,(Waterlow 1995) this reduces to 375 kJ/mol. Additional costs of the polypeptide chain synthesis, neglecting chaperones, is ~11−19 ATP, or 330−660 kJ/mol.(De Visser *et al.* 1992) Amino acid transport and chaperones (which need to be synthesized independently) increase costs further. Under growth conditions where most selection probably occurred historically, very few amino acids are recycled, and thus the specific turnover costs per amino acid in a protein molecule ($C_{s_i} + C_{d_i}$) may easily reach 1500 kJ/mol. However, the amino acid-specific values vary little compared to the protein-specific $k'_{d_i}$ and $A_i$, and thus we use a value of 1500 kJ/mol in **Table 2**. With a typical protein of 400 amino acids, this implies 10$^{-15}$ J s$^{-1}$ of turnover cost per protein molecule, which varies perhaps by 3−4 orders of magnitude, mostly due to $N_{aa_i}$ (protein length) and $C_{s_i}$ (the biosynthetic cost of the amino acids) consistent with the empirically known sequence biases.(Akashi & Gojobori 2002; Swire 2007; Wagner 2005)

The exponential of Equation (1) can be expanded as 1− c$U$ because the values of c$U$ are much smaller than 1. Accordingly, the empirically proposed(Drummond & Wilke 2008) fitness cost constant $c$ can be expressed in terms of fundamental protein turnover parameters, and we argue that $c$ is protein-specific. The PCM fitness function, Equation (7), can be written as:

$$\Phi = \frac{dE_t/dt - \sum_i A_i \exp\left(\frac{\Delta G_i}{RT}\right) k_{d_i} N_{aa_i}(C_{s_i} + C_{d_i})}{dE_t/dt} = 1 - \frac{\sum_i U_i k_{d_i} N_{aa_i}(C_{s_i} + C_{d_i})}{dE_t/dt} \qquad (14)$$

Comparing the exponential-expanded fitness functions $1 - cU$ proposed by Drummond and Wilke(Drummond & Wilke 2008) and Equation (14), the dimensionless protein-specific and effective total cost constants are:

$$c_i = \frac{k_{d_i} N_{aa_i}(C_{s_i} + C_{d_i})}{dE_t/dt}; \quad c = \frac{\sum_i U_i k_{d_i} N_{aa_i}(C_{s_i} + C_{d_i})}{dE_t/dt \, U} \qquad (15)$$

Separation of $U_i$ from its cost constant $c_i$ does not apply in general, as each type of unfolded protein has specific costs, and thus $c$ represents an average cost of handling all misfolded proteins regardless of type. Using the typical values of $k_{d_i} = 10^7$ s$^{-1}$ and $N_{aa_i}(C_{s_i} + C_{d_i}) = 10^{-15}$ J s$^{-1}$ (**Figure 1, Table 2**) gives $10^{-8}$ J s$^{-1}$ for one molecule of protein $i$. When dividing by $dE_t/dt \sim 10^{-11}$ J s$^{-1}$, this gives a cost constant $c_i \sim 1000$. Summing over all misfolded copies (U $\sim 10^{-3}$) gives a correction to the fitness function of the order of unity, in agreement with energy allocated to reproduction and proteome turnover being of the similar magnitudes as total respiration rates of growing cells.(Harold 1987)

A single protein's contribution to fitness is proportional to its relative abundance, all else being equal. If $A_i = 1000$, then $U_i = 10^{-8}$ misfolded copies of this particular protein exist at any time, using the typical parameters of **Figure 1** and **Table 2**, giving a total contribution to fitness of $10^{-5}$. Typically arising, slightly deleterious mutations in typical proteins will affect evolution rates in small populations of the order of N $\sim 10^4$, which probably played a major role in evolution in the wild,(Gillespie 2001; Piganeau & Eyre-Walker 2009) mainly because historic population bottlenecks dominate the apparent effective population size.(Bouzat 2010; Hawks *et al.* 2000; Willis & Orr 1993) The calculation example in **Table 2** gives a fixation probability of $7.9 \cdot 10^{-5}$ for such typical mutations.

However, some proteins are much more systemically important than such a typical protein. The most important contributor to $c_i$ is the degradation rate constant $k_{d_i}$, which varies by many orders of magnitude for different proteins, and to obtain the fitness we need to multiply this constant by $A_i$, or equally, the fold-stability weighted $U_i$. Abundance can span 5−7 orders of magnitude,(Beck *et al.* 2011; Ghaemmaghami *et al.* 2003; Jansen & Gerstein 2000; Milo 2013) whereas protein length $N_{aa_i}$ spans about three orders of magnitude, up to ~30,000 amino acids (titin e.g.), with a reasonably small variance of gamma-distributed protein sizes.(Zhang 2000) PCM theory thus suggests that selection acts both on expression level and protein length, as indeed seen experimentally.(Bloom *et al.* 2006a) In small populations ($N = 10^4$), a typical slightly deleterious mutation (less stable by 5 kJ/mol, or a ten-fold higher turnover rate) in a highly expressed protein ($10^5$ copies) will have essentially no probability of fixation ($< 10^{-20}$, middle right, **Table 2**). Cost selection in such moderate-sized populations can thus explain the relatively slower evolution of abundant proteins.

Large effective populations can also contribute to the E-R anti-correlation: Random mutation-selection dynamics resulting from purifying or compensatory selection of new residues after accepting slightly deleterious mutations occur more frequently in less abundant proteins that have more neutral selection coefficients. In contrast, these dynamics are less important near the steeper fitness optimum of the more optimized, abundant proteins that pose larger costs to the proteome. The relative importance of these two mechanisms depends on the historic effective population size and the population bottlenecks on long evolutionary timescales. One can model such effects by explicit evolution simulations but this is beyond the scope of the present work.

For comparison to experimental values and to further consolidate the theory, it is more convenient to use the fitness function written in Equation (16):

$$\Phi = 1 - \frac{\sum_i A_i N_{aa_i}\left(C_{s_i}+C_{d_i}\right)\ln 2/t_{½i}}{dE_t/dt} \qquad (16)$$

where $t_{½i}$ is the experimental in vivo half live of the protein $i$, which accounts for real cellular life times distinct from biophysical protein stability, e.g. effects of the N-end rule.(Gibbs *et al.* 2014; Mogk *et al.* 2007; Varshavsky 1997) All the properties in Equation (16) are either observable or deducible from the protein's sequence.

**Scaling relations of proteome costs: Mass, metabolism, and eukaryote evolution**

The examples given have centered on yeast as model cell, with $\sum A_i = 10^8$. Eukaryote cells vary greatly in size, the total copy number of proteins, and metabolic respiration rates, and prokaryotes typically feature smaller volumes, protein copy numbers and lower metabolic total respiration rates by 2–3 orders of magnitude.(Milo 2013) The question then emerges how these order-of-magnitude differences affect the proteome turnover and the associated effects described above. Proteins are degraded differently due to specific degrons of their sequences, but the overall rate of protein turnover typically scales with the general activity of the proteasome (except those proteins that are not degraded by the proteasome). Accordingly, a scale factor of proteasome activity α (Equation 6), as modulated by proteasome inhibitors, will be an important control parameter in experimental tests of the theory as well as in efforts to understand protein turnover in relation to cellular energy costs, cell viability, and fitness. Although long-term proteasome inhibition is toxic, mild instantaneous proteasome inhibition should prove a useful tool in testing some of the mechanisms described here.

Additional scaling relations are relevant to discuss. Notably, from Equation (8), any scaling of the metabolic rate by a number *a* characteristic of the organism will not affect the selection coefficient, if the fraction of energy devoted to reproduction is constant, commonly between 0.1 and 0.7 of total respiration costs,(Harold 1987; Hawkins 1991) since the advantage of the mutation with lowered maintenance costs can be considered a perturbation:

$$a\, s_i(M) = a\frac{\Phi_i(M)}{\Phi_i(WT)} - a = \frac{a\Phi_i(M) - a\Phi_i(WT)}{a\Phi_i(WT)} = \frac{adE_m/dt(WT) - adE_m/dt(M)}{adE_t/dt(WT) - adE_m/dt(WT)} = s_i(M) \qquad (17)$$

This relation requires comparison of the mutant and wild type proteins under the same growth conditions.

Based on cell volume and protein copy measurements and associated calculations,(Milo 2013) and using the assumption that a typical protein volume is 10,000 Å$^3$, proteins take up 1−4% of the cell volume of any cell and more importantly, regardless of the cell type, across prokaryotes and eukaryotes, including human cells. From this, we conclude that the total protein copy number $A_i$ scales approximately linearly with cell volume. In contrast, the basal specific metabolic rate of both cells and whole organisms tends to scale with $M^{3/4}$, rather than M (Kleiber's law).(Kleiber 1932, 1947; Savage *et al.* 2007) Size, all-else-being equal, lowers the specific surface area of the organism and thereby increases metabolic efficiency by reducing the mass-weighted thermodynamic force required to maintain the non-equilibrium boundary (reduced heat dispersion per unit of biomass). Size also potentially minimizes average, mass-specific chemical and electric signaling distances within the organism. Such scaling laws of mass and volume and their implication for bioenergetic costs were discussed by Lynch and Marinov.(Lynch & Marinov 2015)

For these reasons, the specific resting metabolism decreases with volume or mass, and equally, with total protein copy number of the organism. Accordingly, size carries an evolutionary advantage of the order of the mass-specific metabolic rate, as explained in detail by Brown, Marquet, and Tape who developed the framework relating mass to evolutionary fitness.(Brown *et al.* 1993) The advantage is of the order of:

$$s(M) = \frac{\Phi(M)}{\Phi(WT)} - 1 = \frac{dE_r/dt(M)}{dE_r/dt(WT)} - 1 = \frac{-aM^{\frac{3}{4}}(M)}{-aM^{\frac{3}{4}}(WT)} \sim \left(\frac{M(M)}{M(WT)}\right)^{3/4} \qquad (18)$$

However, as pointed out by Brown et al.,(Brown *et al.* 1993) whereas ecological life-history variables, notably foraging efficiency, favors large organisms, the time of reproduction is favorable for smaller organisms and scales with $M^{-1/4}$. Thus, organism size has an evolutionary optimum with respect to both energy and time, which is distinct for different taxa due to the different life-history variables and associated scaling parameters.(Brown *et al.* 1993) A yeast mutant with a larger size of 1%, all-else being equal, would thus be predicted by PCM theory to have a selective advantage of $(1.01/1)^{3/4} = 0.007$ if all the saved energy is spent on reproduction. This energy is clearly enough to enforce positive selection at all relevant population sizes from $10^2$ to $10^7$, including early population bottlenecks (**Figure 2**).

Combining the ansatz of PCM theory (that fitness scales with the energy left for reproduction per time unit after subtracting maintenance costs) with Kleiber's law leads to several potentially important explanations for size advantage relevant to emergence of life in general and eukaryotes in particular. A central weakness of endosymbiont theory, not mentioned by the otherwise important reviews on this topic,(Gray *et al.* 1999; Lane 2011) is the problem of evolutionary advantage *immediately* after the symbiosis event. The argument goes as following: At the very beginning, the actual process of symbiosis must have had immediate costs of intrusion and aligning the cellular machineries, and must thus also have provided immediate selective advantages in competition will non-symbiotic cells. According to PCM theory, fitness scales with energy left for reproduction, and thus the immediate total maintenance costs must have reduced.

Imagine a simple doubling of the cell size by a unification event. All else being equal, the new organism would carry the double amount of proteins, the double volume, the double mass, and would require the double amount of energy to reproduce these cell constituents, giving the same fitness as the competing non-symbiont cells, but then reduced by the costs of the endosymbiosis event itself. However, the immediate advantage offered by reducing the specific surface area of the ancestral eukaryote cell would reduce the basal metabolic maintenance rate. The saved energy could then be immediately converted into a larger fraction of the total energy budget being devoted to the proteome of larger cells and organisms, thus compensating the cost of the actual symbiosis event. If this is correct, endosymbiosis will be successful only when and if the mass-specific metabolic rate saved by mass increase outweighs the energy costs of the symbiosis event itself.

**Evidence for PCM during evolution**

Support for the theory of proteome cost minimization is evident at many levels and time scales of evolution, with some examples summarized in **Table 3**. The following section discusses some of these facts briefly.

*Major evolutionary events mainly represented bioenergetic advantages.* During the longest and earliest timescales where much of the primary cellular biochemistry evolved, unicellular growth conditions provided the context for the evolutionary innovation both in terms of respiration and photosynthesis.(Blankenship 1992; Sousa *et al.* 2013) Most of the important biochemical pathways being at least qualitatively evolved at the point when eukaryotes had formed.(McGuinness 2010; Nisbet & Sleep 2001) Early qualitative innovations such as the electron transport chain, fatty acid and amino acid metabolism, and photosynthesis indicate the primary importance of obtaining and maintaining the bioenergy production,(Sousa *et al.* 2013) a tendency further documented by the rise of eukaryotes whose advantages largely related to energy efficiency by outsourcing and optimizing energy production as argued above and elsewhere.(Gray *et al.* 1999; Lane 2011; Margulis 1968, 1975)

*Energy surplus determines growth of microorganisms.* For unicellular organisms, the cell cycle determining the decision to grow (and thus contribute to population fitness) is largely based on an assessment of available energy:(Cai & Tu 2012) Thus, budding yeast grows during the G1 phase until the nutrient level determines whether it commits to reproduction and enters the DNA biosynthesis S phase and subsequent mitosis, or if cell growth is arrested due to low resources.(Cai & Tu 2012)

*Protein turnover is very expensive.* Protein turnover is typically the most or second-most expensive process in cells: At one extreme, protein synthesis may account for 3/4 of all energy spent in growing microorganisms.(Harold 1987) In humans, protein synthesis typically requires 20 kJ/kg body mass, or 20% of the basal metabolic rate to produce typically 300 g of protein per day.(Reeds *et al.* 1985; Waterlow 1995) This number does not include regulation and degradation costs, RNA synthesis, and uncertain costs relating to nitrogen metabolism, reuse, transport, or synthesis of amino acids, which together are substantial.(Hawkins 1991; Reeds *et al.* 1985) In mammals, protein degradation may cost 10−20% of total energy spent.(Fraser & Rogers 2007; Hawkins 1991) Ubiquitin requires ATP to bind proteins targeted for degradation, and the lysosome and calcium-dependent proteases require ATP for active calcium and proton transport.(Hawkins 1991) These various features render protein turnover (synthesis and degradation) the most or second-most (next to ion pumping) energy-consuming process even in mammals.

*Life uses cheap amino acids.* The synthetic costs of the 20 amino acids vary roughly from the order of ~10 (Glu, Ala, Gly, etc.) to ~75 (Trp) phosphate bonds.(Akashi & Gojobori 2002; Heizer *et al.* 2011) Biosynthetic costs explain some of the amino acid bias in sequences not due to translational efficiency and other effects(Akashi 2003; Akashi & Gojobori 2002; Craig & Weber 1998) and can affect the rate of evolution.(Barton *et al.* 2010) Selection towards cheaper amino acids or smaller proteins can reduce total energy expenditure substantially, by an estimated 0.1% per ~4 expensive amino acids.(Akashi & Gojobori 2002) A general evolutionary preference for synthetically cheap amino acids was first suggested (for aromatic residues in *E. coli*)(Lobry & Gautier 1994) and later demonstrated(Akashi & Gojobori 2002) and confirmed by others(Heizer *et al.* 2006; Wagner 2005) in prokaryotes, where cheaper amino acids tend to be used more in highly expressed proteins across functional classes, with similar observations seen for yeast.(Raiford *et al.* 2008) These findings have been confirmed in many cases(Garat & Musto 2000; Heizer *et al.* 2011; Kahali *et al.* 2007; Raiford *et al.* 2008) including mammals.(Heizer *et al.* 2011) Biosynthetic cost minimization as an evolutionary driver was identified first in certain bacteria(Akashi & Gojobori 2002; Schaber *et al.* 2005) and later in all domains of life(Swire 2007). Cys is apparently not significantly selected for cost,(Swire 2007) perhaps relating to its unique involvement in highly conserved cystine bridges and metal sites.

*Prokaryote streamlining.* The fact that prokaryotes have maintained their general morphology until today whereas Eukarya is represented by rich morphological diversity reflects the existence of some selection pressure that kept prokaryotes simple but afforded major degrees of freedom to Eukarya. The well-known intense streamlining of the small efficient prokaryote genomes has led to the formulation of the so-called streamlining theory of microbial evolution,(Giovannoni *et al.* 2014; Lynch 2006) which argues that streamlining towards small efficient genomes have been an ongoing selection pressure of prokaryote evolution. The theory and observations fit very well to the predictions of PCM theory, as any energy saved on maintaining genomes can be diverted towards reproducing the prokaryote cell, although the advantage can be both in terms of energy and time.

*Highly expressed proteins are more streamlined.* Highly expressed genes tend to code for smaller proteins(Jansen & Gerstein 2000) with less introns,(Urrutia & Hurst 2003) in support of selection pressure towards minimizing proteome handling costs. Selection against mistranslation can also be understood as selection against biosynthetic cost because translational efficiency is effectively a way to minimize the cost of GTP-dependent "proofreading" and other machinery operating on mistranslated gene products.(Ikemura 1985) Adddional support for the selection on highly abundant proteins directly relating to turnover costs is the well-known relationship between expression levels and protein half life.(Belle *et al.* 2006)

*Some proteomes change adaptively in nutrient-, N- and S-restricted habitats.* Sparse, potentially growth-limiting, essential amino acid precursors can selectively change the composition of proteomes.(Baudouin-Cornu *et al.* 2001) N-restricted marine microorganisms(Grzymski & Dussaq 2012) and S-restricted cyanobacteria(Mazel & Marliére 1989) both undergo changes in their proteomes in response to the reduced availability of these elements. In some spiders,(Craig *et al.* 2000) nutrient levels and diet directly affect the composition of the proteome's use of amino acids. In the case of N-restricted marine microorganisms, the proteome changes were directly related to selective advantages of biosynthetic cost minimization for highly expressed proteins.(Grzymski & Dussaq 2012)

*Unstable proteins reduce cell growth.* Support for the PCM theory also comes from studies that compare the biophysical properties of overexpressed wild-type and mutant proteins directly. Destabilizing mutants of lacZ in *E. coli* reduce cell growth to a similar extent as wild-type protein expressed at the same level, arguing for quantity (expression levels subject to turnover) as the cause of toxicity rather than qualitative features of the protein variants.(Plata *et al.* 2010) An implication of this is that reduced cell viability in assays of overexpressed misfolding proteins, often used as models of neorodegenerative disease, may in fact reflect energy deficits as described by PCM theory. If so, misfolded proteins are generally not toxic by a specific mode of action (such as membrane pore formation or seeding of misfolding leading to loss of function) but rather because of the ATP costs.(Kepp 2019)

**Trading function for cost**

Classical Darwinian evolution considers the struggle and selection for optimal function the primary mode of evolution.(Hurst 2009; Richmond 1970) This aspect of Darwinism has dominated biochemical views of enzymes as perfectly optimized proficient catalysts that accelerate chemical reactions by orders of magnitude, implying that evolution strives towards maximal proficiency *per se*(Cannon *et al.* 1996; Radzicka & Wolfenden 1995; Zhang & Houk 2005) However, as already mentioned, non-function universal selection pressures operate more generally on larger timescales of evolution,(Bloom & Adami 2003, 2004; Drummond *et al.* 2005; Hurst & Smith 1999; Lobkovsky *et al.* 2010; Wylie & Shakhnovich 2011) and actual comparison of enzyme kinetic parameters shows that many enzymes are distinctly suboptimal, most likely because of evolutionary and biophysical constraints.(Bar-Even *et al.* 2011)

A standard view is that proteins have evolved to use their excess fold free energy to optimize the active sites for function, the most notable example being pre-organized active sites with electrostatic fields favoring the free energy of the transition states, to increase $k_{cat}/K_M$.(Adamczyk *et al.* 2011; Cannon *et al.* 1996; Fuller *et al.* 2019; Morgenstern *et al.* 2017; Warshel 1998) Although not directly pointed out by Warshel and co-workers, this mechanism contributes to making proteins marginally stable because, all-else-being equal, any potential excess fold free energy has been diverted into optimizing the electrostatic field of the folded structure to reduce the transition state's free energy and thereby increase catalytic proficiency. The mechanism also largely explains the widely observed stability-function trade-offs in protein engineering(Tokuriki *et al.* 2008). Correspondingly, "designability" or evolvability tend to follow from the fact that outside the biological context, protein engineering can remove many constraints and thereby additionally optimize function of a high-stability starting protein.(Bloom *et al.* 2006b; Tokuriki & Tawfik 2009) This is particularly relevant in the context of "directed evolution", i.e. the intended human evolution of new improved protein mutants, as originally proposed by Francis and Hansche in 1972(Francis & Hansche 1972) employing yeast cells with short generation times in static environments where selection pressure can be effectively controlled, and later demonstrated also in *E. coli* K12 by Barry Hall in 1981.(Hall 1981)

PCM theory argues that even functional proficiency often evolved conditionally on cost. To appreciate this, we consider the requirement of a certain total substrate turnover of each enzyme per time unit to maintain homeostasis. The proficiency of function is for enzymes typically defined by $k_{cat}$, measuring how many substrate molecules convert into product per time unit per enzyme molecule. At steady-state, both the maximum turnover ($V_{max}$) and the turnover at low substrate concentration is proportional to the total enzyme concentration [E] and $k_{cat}$.(English *et al.* 2005; Northrop 1998)

Now consider a typical arising mutation in an enzyme *i* required to make a product at a certain rate, i.e. *dPi/dt*. Because the protein is evolutionarily optimized (but not necessarily optimal), mutations will tend on average to be hypomorphic and reduce the turnover constant $k_{cat,i}$ but with a broad scatter and many nearly neutral effects with a random chance of fixation. If the mutation reduces $k_{cat,i}$ substantially, e.g. by modifying the active site, the substrate turnover will be greatly reduced, and the organism will need to increase the local enzyme concentration [E] by expressing more enzyme per time unit to maintain a comparable substrate turnover (compensatory expression), thereby increasing $A_i$. More specifically, the rate of product formed by enzyme *i* under Michaelis-Menten kinetics is:(Cannon *et al.* 1996; Northrop 1998)

$$\frac{dP_i}{dt} = A_i \, k_{cat,i} \frac{[S]}{K_{M,i}+[S]} \tag{19}$$

Equation (19) represents the standard equation multiplied on both sides by the cell volume to convert from concentrations to absolute copy numbers. For simplicity, we can ignore the last term and assume zero order kinetics in [S], which represent selection of the enzyme for maximum rate at saturated substrate concentration when [S] is much larger than the Michaelis constant $K_{M,i}$. The cost of maintaining the enzyme is:

$$dE_{m,i}/dt = A_i k'_{d_i} N_{aa_i}(C_{s_i} + C_{d_i}) \tag{20}$$

Accordingly, the specific cell-wide cost of maintaining steady state produced concentration of $P_i$ is:

$$dE_{m,i}/dP_i = \frac{A_i k'_{d_i} N_{aa_i}(C_{s_i}+C_{d_i})}{A_i k_{cat,i}} = \frac{k'_{d_i}}{k_{cat,i}} N_{aa_i}(C_{s_i} + C_{d_i}) \tag{21}$$

If measured in concentrations instead, the cost scales with the volume of the cell $V_{cell}$ to which the steady state applies. We have ignored the costs associated with producing the substrate and transporting the substrate and products, which can easily be included into the model.

Equation (21) predicts that the ratio of the two time constants for turnover of the enzyme and turnover of the substrate together define the cost of producing $P_i$ at steady state. The two time constants are in units of s$^{-1}$, and $N_{aa_i}(C_{s_i} + C_{d_i})$ is of the order of $10^{-15}$ J for a typical protein. Considering again a typical arising mutation, even if $k'_{d_i}$ is not increased (which it typically is), a reduction in $k_{cat,i}$ of a typical hypomorphic mutation will require compensatory expression of the enzyme, increasing $A_i$ to maintain the rate of production of $P_i$, Equation (19). This increase in $A_i$ will then increase the total cost of obtaining the product with the same factor(Equation 20). Equation (21) summarizes this cost-function relationship since $k_{cat,i}$ and $A_i$ are inversely related if homeostasis in $P_i$ is required. If compensatory expression is 100%, a ten-fold reduction in the enzyme's $k_{cat,i}$ requires a ten-fold increase in the enzyme's expression, and the specific and total costs of producing $P_i$ increases ten-fold.

Accordingly, even mutations that only impair function also increase the proteome costs: A 10-fold increase in $k'_{d_i}$ (loss of kinetic stability, misfolding) or decrease in $k_{cat,i}$ will have approximately the same 10-fold increase in cellular costs, according to Equation (21), ignoring the mutation-induced changes in the amino-acid synthesis and degradation costs. If required, the assumption of 100% compensatory expression can easily be modified by a scale factor between 0 and 1 in the equations above. Evidence for compensatory expression is well-known, a dramatic example being homozygous sickle cell disease (**Table 3**), where dysfunctional, instable hemoglobin mutants cause a doubling of protein turnover and degradation in patients and a 20% in total resting metabolism.(Badaloo *et al.* 1989) Considerations of loss and gain of function mutations associated with other diseases may be viewed in this light.(Kepp 2015, 2019)

Because of the above considerations, we expect a function-cost tradeoff acting during evolution of many proteins. We obtain the important possibility that *the main advantage of a mutant may not be a functional improvement of the protein per se, but a reduction its cost per unit of function in the simplest case the ratio* $k'_{d_i}/k_{cat,i}$. Co-optimization of cost vs. function is fundamental to many optimization processes and follows the basic principle that if several inputs are available at different functionality and price, the optimal system uses the input whose cost per unit of function is lowest.

Such systems will tend to use less functional input if its cheaper price outweighs the loss of function. This suggests that at least some of the widely observed inverse relationships between function and stability(Bonet *et al.* 2018; Du *et al.* 2018; Tokuriki *et al.* 2008) in reality reflect a cost-function tradeoff as summarized by Equation (21). The laboratory can change selection pressures drastically away from those in the wild, notably in the form of "directed evolution".(Francis & Hansche 1972; Hall 1981) In nature however, the situation is more complicated, because the stability affects the proteome costs and thus fitness. Newly arising mutations may impair both stability and function, but both have a direct negative fitness effect in terms of cost.

The theory thus predicts that highly abundant proteins, because they are more cost-selected, are more likely to display suboptimal functionality, all else being equal (after adjusting for other correlating variables such as size). The tradeoffs will be habitat- and strategy-dependent, and the preferential use of very functional but expensive input may be restricted to high-nutrient habitats and growth media.

**Time or energy?**

We expect that variations in the habitat's selection pressure should affect the proteome function-cost tradeoffs. This should be evident when comparing organisms adapted to different environments. The most obvious biophysical properties of the habitat are time, energy, and temperature, which all enter directly in the model, Equation (12). Time enters via the central ansatz of the theory, that "*fitness is proportional to the energy per time unit available for reproduction after subtracting maintenance costs*", i.e. Equation (6). $\Phi = dE_r/dt = dE_t/dt - dE_m/dt$. Fitness scales inversely with the time step dt required for directing a unit of surplus energy sufficient to complete a reproductive event. Temperature enters as a modifier of the protein stability's role in the turnover $\Delta G_i/RT$. Accordingly, all of these biophysical properties enter as selection pressures in the model and are expected to shape evolution, as described below. Obviously, selection pressures on size, cost, and speed correlate to some extend, as small genomes are faster and cheaper to reproduce. Furthermore, translational efficiency could improve both the speed and cost of reproduction if the reduced cost of turnover of the fewer misfolded protein copies outweigh the cost of quality control.

The theory's describing parameters can be determined experimentally, and one can test whether one or another biophysical parameter is restricting growth via competitive growth assays with variable space and energy restrictions. One can consider r- and k-strategies as extreme outcomes specialization to niches of a habitat that is heterogeneous in the distribution of space, time, and energy. The following mechanisms can be postulated: 1) If energy and space is plenty and no new functions evolve, selection will act mainly on reproduction speed (survival of the fastest). 2) If energy is limited, the most energy efficient organisms a likely to prevail (survival of the cheapest). 3) If space is limiting growth, expensive or time-consuming innovations that minimize cell volume and increase the colony's ability to adapt to a distinct spatial geometry of the habitat may be successful, e.g. layered growth.

Mixtures of strategies and selection pressures probably occur in specific cases. One recent study that casts light on this is a study of pathways choices among different sequenced organisms(Du *et al.* 2018). The study found that different organisms select specific choices of precursor pathways based on both metabolic cost and synthetic efficiency. Cost selection turns on in organisms evolving in energy-poor habitats, whereas in energy-rich habitats, the default selection mode is time. Again, there are correlations between time and energy advantages. Notably, the synthesis time of more expensive amino acids is likely longer as it requires more phosphate bonds diverted during synthesis. The cost of handling misfolded proteins can limit growth substantially, as seen in a case of ~3% growth rate reduction in yeast upon folding-stability-impaired mutants of only one protein (YFP).(Geiler-Samerotte *et al.* 2011)

The shift in selection pressure from time to energy can also explain the important phenomenon of overflow metabolism, the tendency of using more expensive, but faster fermentation rather than respiration during growth.(Basan *et al.* 2015) Based on the theory described above, we predict that microorganisms shift to fermentation and selection occurs mainly on fermentation in rich habitats and growth media, because time is the main selection pressure, whereas in poor habitats, respiration becomes favored and selected upon because energy is restrictive. This mechanism largely explains the microbial behavior in many growth assays, but also the Warburg effect of cancer cells.(Basan *et al.* 2015) Cancer cells are remarkable by being under selection both for time, space, and energy in direct competition with each other against the selection pressure of the body's immune system and spatial and nutritional constraints. For this reason, cancer cells tend to use cheaper amino acids,(Zhang *et al.* 2018) in accordance with PCM theory. This insight may have consequences for cancer research although this is beyond the scope of the present paper.

**Temperature, thermostable proteins, and thermophilic organisms**

As mentioned above, the habitat temperature also imposes a selection pressure on evolution according to the PCM theory, because it directly modifies protein stability $\Delta G_i/RT$ and thereby, the fitness function, Equation (11). To appreciate this, we used a sign convention of negative $\Delta G_i$ for a stable protein, and the $\Delta G_i$ is the optimal stability of the protein at its temperature of operation (sometimes called T*), typically reflecting to some extent the organism's experienced extrema temperatures in the relevant habitat.(Robertson & Murphy 1997) The protein has been optimized to display its maximal stability at this T*, with $\Delta G_i$ typically harmonic in the temperature, and increasing or decreasing the temperature away from T* will thus increase the number of misfolded proteins $U_i$ and increase the associated turnover costs, thereby reducing fitness, Equation (11).(Robertson & Murphy 1997)

Using the theory, we can better understand adaptation of proteomes to hot or cold environments (thermophiles and psychrophiles, respectively).(Fu *et al.* 2010; Li *et al.* 2005; Luke *et al.* 2007; Mozo-Villiarías & Querol 2006) Adaptations to a warmer habitat is largely expected to be a question of optimizing the proteome's copy-number-weighted median protein T* (the most representative T* of the proteome of the cell) towards the T of the habitat, in order to minimize the average copy number of misfolded protein copies in the cell at any given time, again to minimize proteome costs and maximize energy available for reproduction. Many studies of thermophilic proteins and thermophilic adaptation may be seen in this light, without going into further detail, as this is a large and complex topic,(Sawle & Ghosh 2011; Tekaia *et al.* 2002; Venev & Zeldovich 2018) but the essential implications should be clear. In particular, thermophilic organisms are predicted to adjust protein thermostability mainly for the most abundant and quickly turned-over proteins that pose the largest economical cost to the proteome.

**PCM, aging, and neurodegenerative diseases**

Proteome cost minimization has been argued to explain a substantial part of the evolution on longer evolutionary timescales, producing clear biases in the use of amino acids and explaining the E-R anti-correlation by slowing the probability of fixating new mutations in abundant, expensive proteins, and giving rise to important cost-function trade-offs. The evolution that shaped these relations mainly occurred in single-cell organisms, and it is thus of interest to consider whether the theory has implications also for evolution of higher organisms and in particular the evolution of aging.

All higher organisms use oxidative phosphorylation as the most effective energy-producing process, using the $O_2$ of the planet's atmosphere produced by the photosynthetic organisms as primary electron acceptor. The free radical theory of aging argues that aging arises from the incurred damage due to the activity reducing $O_2$ to water, as the radical side products of the respiratory chain leads to a consistent mutagenic pressure that needs to be countered by DNA repair and antioxidant defenses.(Harman 2003; Speakman *et al.* 2002)

Different higher organisms have evolved different trade-offs between life history variables relating mainly to the generation time.(Kirkwood 2011; Kirkwood & Rose 1991; Shanley & Kirkwood 2000) Shorter lifespan implies specialization towards shorter generation time, which again implies less energy invested in maintenance of the proteome. Based on the discussion above, this specialization emphasizes time over energy. Each strategy probably involves an aging program to "dispose the soma" after reproduction to make space for the next generation, although this remains debated.(Speakman *et al.* 2002; Westendorp & Kirkwood 1998) Sexual reproduction, which is a major advantage in terms of genetic variation, evolutionary and adaptive potential, and robustness to habitat perturbations such as climate change, emphasizes these strategies.(Kirkwood 2001; Kirkwood & Austad 2000) Aging may thus be a direct consequence of the reproductive strategy.

Some organisms that specialized towards strategies of long lifespan (i.e. *r*- vs. *k*-strategists) also diversified towards complex lifestyles with capacity for technology transfer between generations, e.g. cetaceans and apes. Compared to mammals, rodents on average have shorter generation times, lifespans, larger litter size, and have traded lifespan for fecundity.(Speakman *et al.* 2002; Wensink *et al.* 2012) In long-living organisms, proteome misfolding may cause death, perhaps because PCM can no longer be afforded beyond what was evolutionarily beneficial, in terms of fitness. It is reasonable to argue that the aging program of long-living mammals largely reflect the (active or passive) giving up of the maintenance of the proteostatic machinery to enable the rise of the next generation.(Hipkiss 2017; Taylor & Dillin 2011)

The discussion is greatly facilitated by considering superoxide dismutase 1 (SOD1). SOD1 is one of the most abundant proteins in primates and $A_i$ can reach 100,000 copies per cell,(Dasmeh & Kepp 2017) it is the central antioxidant defense protein of the mitochondria thus directly linking energy and aging,(Perry *et al.* 2010) it is one of the few proteins known to directly extend lifespan upon induction,(Landis & Tower 2005; Tolmasoff *et al.* 1980) and one of the few genes of great apes known to have undergone non-synonymous positive selection.(Dasmeh & Kepp 2017; Fukuhara *et al.* 2002) Deposits of misfolded SOD1 is a hallmark of age-triggered amyotrophic lateral sclerosis.(Valentine *et al.* 2005) The tendency towards aggregation and misfolding of natural human SOD1 variants correlates with their pathogenicity,(Kepp 2015; Lindberg *et al.* 2005; Wang *et al.* 2008) and wild type overexpression by itself is enough to trigger disease.(Wang *et al.* 2009) Recent amino acid substitutions in SOD1 of great apes correlate with longer life span and tend to increase the net charge and stability of SOD1, thus increasing the thermodynamic and kinetic stability of the protein, ($k_d$ and $\Delta G_i$).(Dasmeh & Kepp 2017) Via its abundance and functional importance, any impairment of SOD1 either in terms of function or stability will produce comparatively very large PCM costs. The combination of the features summarized above strongly argue for a relationship between PCM, evolution of aging, and age-triggered neurodegenerative diseases, as recently emphasized.(Kepp 2019)

The differences in life history variables between rodents and primates, and in particular great apes, produce serious challenges for the use of murine models of neurodegenerative disease, which all have age as their main risk factor. According to the PCM theory, neurodegenerative diseases are caused by the increased energy spent on maintaining the proteome of old humans, which leaves less energy available for neuron and motor neuron function. Protein turnover and neuron signaling costs perhaps 20−25% and 50% of the brains energy budget,(Attwell & Laughlin 2001; Hawkins 1991; Raichle & Gusnard 2002) respectively, and as age advances, the supply of energy may reach the point where it is no longer sufficient to satisfy the increasing maintenance costs of the proteome.(Kepp 2019) Familial inherited mutations that tend to produce more aggregation-prone protein will increase turnover costs per time units according to PCM theory and will accordingly also accelerate the time at which the supplied energy no longer satisfies the needs of synaptic transmission, leading to memory loss with earlier clinical age of onset.(Kepp 2019) In sporadic forms of diseases and normal aging, the protein deposits grow monotonically and become critical in people with accumulated genetic and life-style imbalances that cause elevated proteome costs or impaired energy supply, e.g. diabetes and obseity.(Diaz 2009; Ott *et al.* 1996) Applying the PCM theory and these observations to the field of protein misfolding diseases suffered by millions of people worldwide thus seems to be high priority for the future.

**Conclusions**

Darwin's theory of evolution emphasized "survival of the fittest", where the "fit" represented optimal functional proficiency. This concept has dominated the thinking of the field, including the biochemical view of enzymes as optimally proficient for their catalytic reaction.(Radzicka & Wolfenden 1995; Zhang & Houk 2005) Proteomic data have shown that most effects on the speed of evolution act via non-functional, universal selection pressures.(Drummond *et al.* 2006; Pál *et al.* 2001, 2006) The main outstanding challenge in evolution is arguably to provide a predictive quantitative theory that captures these universal selection pressures and predicts real evolutionary histories, including the relative magnitude of drift and selection in specific cases, the nature of the selection pressures, and how it acts upon a population via the individual, the cell, the protein, and the gene.

This paper has reviewed the theory that the universal background selection pressure of life is minimization of the ATP cost of an organism's proteome ("survival of the cheapest"). The magnitude and variations of the fundamental parameters show that most of the proteome cost selection acts via the ratio $A_i / t_{½}$, i.e. the abundance to half life ratio of the protein. This selection combines with the selection for functional proficiency, typically in a cost-function trade-off between being "fit" and "cheap". The data in **Table 2** suggest that cost selection occurred both during the earliest period of prokaryote evolution, during the rise of eukaryotes, particularly explaining the immediate advantages of the larger eukaryote cells due to reduced mass-specific metabolic costs, and during the long periods of relatively uneventful nearly neutral evolution that maintains nearly constant molecular clocks of many phylogenies.

The theory has several implications that could be explored further, e.g. for stability-function and time-energy tradeoffs, the Warburg effect, thermophile evolution, and human neurodegenerative diseases. One implication of the theory is that nature has not generally evolved the most proficient enzymes, in terms of turnover numbers ($k_{cat}/K_M$), but the *lowest cost of substrate turnover*, as given by the ratio of Equation (21). The theory thus predicts that most proteins may be engineered to obtain higher functional proficiency but that this will typically come with an associated increased total cost of the protein pool (either by abundance or specific costs per protein), which may however be less of an issue in the laboratory. The breakdown of this cost-function tradeoff may be a central reason why directed evolution and protein engineering strategies that aim to enhance protein performance even for natural functions are successful at all.

**Financial support**

This research received no specific grant from any funding agency, commercial or not-for-profit sectors.

**Conflicts of Interest declaration**

The author declares that he has no conflict of interest associated with this work.

**Table 1. Important correlators of the evolution rate and size of proteins.**

| Features that slow evolution | Effect | Name |
|---|---|---|
| Functional active sites | Sites directly involved in e.g. recognition, substrate binding, and catalysis are highly conserved.(Blundell & Wood 1975; Casari *et al.* 1995) | Function-rate (F-R) anti-correlation (sequence conservation) |
| High expression | Highly expressed proteins (measured by mRNA levels) evolve more slowly.(Drummond *et al.* 2005; Pál *et al.* 2001) | Expression-rate (E-R) anti-correlation |
| Intracellular location | Intracellular proteins evolve more slowly than extracellular proteins.(Julenius & Pedersen 2006; Winter *et al.* 2004) | Secretion-rate (S-R) correlation |
| Buried amino acid sites | Interior sites evolve more slowly than solvent-exposed sites.(Goldman *et al.* 1998; Overington *et al.* 1992; Ramsey *et al.* 2011) | Buried-rate (B-R) anti-correlation |
| Small size | Smaller proteins, all-else being equal, evolve slowly.(Bloom *et al.* 2006a) Small proteins are less evolvable due to larger functional density.(Zuckerkandl 1976) | Size-rate (S-R) correlation; functional density |
| Small contact density / fraction of buried sites | Proteins with smaller fractions of buried sites or contact density evolve slowly (both strongly correlated with size).(Bloom *et al.* 2006a, 2006b) | Size-rate (S-R) correlation |

**Table 2. Effect of arising mutants in a haploid organism ($N_{aa_i}$ = 400; $dE_t/dt$ = 3·10⁻¹¹ J s⁻¹).**[a]

| | $A_i$ | $k'_{d_i}$(WT) $k'_{d_i}$(M) | $N_{aa_i}(C_{s_i} + C_{d_i})$ | $dE_m/dt$ (WT) | $dE_m/dt$ (M) | $s_i$ (M) | $P_{fix}$ N = 10⁶ | $P_{fix}$ N = 10⁴ |
|---|---|---|---|---|---|---|---|---|
| Slightly deleterious mutant that increase $k'_{d_i}$ or $A_i$ 10-fold (e.g. from 60 to 54 kJ/mol stability at 37°C) | | | | | | | | |
| Total proteome | 10⁸ | 10⁻⁴ s⁻¹ 10⁻³ s⁻¹ | 10⁻¹⁵ J/protein | 10⁻¹¹ J s⁻¹ | 10⁻¹⁰ J s⁻¹ | Cell dies (proteome destabilization corresponds to T = 72 °C) | | |
| Typical protein | 10³ | 10⁻⁴ s⁻¹ 10⁻³ s⁻¹ | 10⁻¹⁵ J/protein | 10⁻¹⁶ J s⁻¹ | 10⁻¹⁵ J s⁻¹ | −4.5·10⁻⁵ | <10⁻²⁰ | **7.9·10⁻⁵** |
| Abundant protein | 10⁵ | 10⁻⁴ s⁻¹ 10⁻³ s⁻¹ | 10⁻¹⁵ J/protein | 10⁻¹⁴ J s⁻¹ | 10⁻¹³ J s⁻¹ | −4.5·10⁻³ | <10⁻²⁰ | **<10⁻²⁰** |
| Shortlived protein | 10³ | 10⁻² s⁻¹ 10⁻¹ s⁻¹ | 10⁻¹⁵ J/protein | 10⁻¹⁴ J s⁻¹ | 10⁻¹³ J s⁻¹ | −4.5·10⁻³ | <10⁻²⁰ | **<10⁻²⁰** |
| Positive selection of slightly beneficial mutant that decreases $k'_{d_i}$ 10-fold | | | | | | | | |
| Typical protein | 10³ | 10⁻⁴ s⁻¹ 10⁻⁵ s⁻¹ | 10⁻¹⁵ J/protein | 10⁻¹⁴ J s⁻¹ | 10⁻¹⁵ J s⁻¹ | 4.5·10⁻⁶ | 4.6·10⁻⁶ | 1.0·10⁻⁴ |
| Abundant protein | 10⁵ | 10⁻⁴ s⁻¹ 10⁻⁵ s⁻¹ | 10⁻¹⁵ J/protein | 10⁻¹⁶ J s⁻¹ | 10⁻¹⁷ J s⁻¹ | 4.5·10⁻⁴ | 4.5·10⁻⁴ | 4.5·10⁻⁴ |
| Neutral evolution (same for all protein properties, only depends on N) | | | | | | | 10⁻⁶ | 10⁻⁴ |

[a]: WT = wild type value of property. M = Mutant value of property.

**Table 3. Events and facts supporting the PCM theory.**

| Observation | Interpretation |
|---|---|
| Protein turnover is very expensive, in particular in growing microorganisms. | The cost of handling the proteome is the most or second-most costly process in many cells(Fraser & Rogers 2007; Reeds *et al.* 1985; Waterlow 1995), and can dominate total energy costs in growing microorganisms.(Harold 1987) |
| Energy surplus determines growth of microorganisms. | In the yeast cell cycle, available energy determines whether the cell commits to reproduction or if growth is arrested.(Cai & Tu 2012) |
| All kingdoms of life favor synthetically cheap amino acids.(Akashi & Gojobori 2002; Garat & Musto 2000; Heizer *et al.* 2011; Kahali *et al.* 2007; Raiford *et al.* 2008; Schaber *et al.* 2005; Swire 2007) | Cheaper amino acids confer a selective advantage by lowering overall protein synthesis costs of the organism. |
| Cheap amino acids are more used in highly expressed proteins.(Ikemura 1985; Seligmann 2003; Swire 2007; Wagner 2005) | Abundant proteins contribute more to total fitness, making cheaper amino acids are particularly advantageous, supporting a relation to both abundance and protein-specific costs. |
| Extracellular proteins use cheaper amino acids.(Smith & Chapman 2010) | Extracellular proteins are not recycled and thus, their net amino acid costs are larger per protein copy, this seems to have been selected against by favoring cheap extracellular amino acid use. |
| Highly expressed proteins tend to be smaller.(Bloom *et al.* 2006a; Ikemura 1985) | Seen in 27 of 31 functional categories of yeast, with 12 classes significant.(Bloom *et al.* 2006a; Ikemura 1985) Length is inversely related to gel-derived protein abundance.(Futcher *et al.* 1999) |
| Cheap amino acids are used in large proteins.(Ikemura 1985; Seligmann 2003) | All-else-being-equal, larger proteins constitute larger turnover costs (weighted by their copy numbers) and thus are more relevant for overall PCM. |
| Large proteins tend to be more stable. | Large proteins tend to be more stable (significant but with large variation).(Sawle & Ghosh 2011) |
| Streamlining theory (the theory that selection favors minimal cell | The intense streamlining of prokaryote genomes(Giovannoni *et al.* 2014; Lynch 2006) reflects selection pressure either via energy, time, or both, and is thus explained by PCM theory |
| complexity).(Giovannoni *et al.* 2014) | |
| Parasites feature reductive evolution on biosynthesis and metabolism.(Loftus *et al.* 2005) | Parasites mainly get their energy and nutrients from the host and thus can increase fitness by adaptive loss of biosynthetic and metabolic pathways. |
| Genes with less intronic DNA more highly expressed.(Urrutia & Hurst 2003) | Less introns probably reduce the cost of protein translation. |
| Protein synthesis efficiency affects the age-dependent growth of blue mussels.(Hawkins *et al.* 1986) | Genetic differences in protein turnover efficiency contribute to fitness in some organisms. |
| Misfolded proteins can reduce yeast fitness/growth by 3.2%.(Geiler-Samerotte *et al.* 2011) | Misfolded proteins impose a cost on the proteome in proportion to the steady state level of misfolded copies and their turnover rate (Equation 9). |
| Evidence for shifts towards economical biosynthetic precursor pathways.(Du *et al.* 2018) | Many organisms have different choice of precursor synthesis pathways, which seem to reflect cost minimization. |
| The endosymbiosis leading to eukaryotes was an energy optimization event.(Lane 2011; Margulis 1975) | The specialized energy production in mitochondria and the associated genomic asymmetry gave rise to enormous expansions and innovations typical of Eukarya.(Lane 2011) |
| Overflow metabolism (Warburg effect in cancer cells).(Basan *et al.* 2015) | The shift in selection pressure from time to energy explains overflow metabolism, because fermentation is faster but respiration is cheaper. |
| Cancer cells use cheaper amino acids.(Zhang *et al.* 2018) | Cancer cells are under selection for both time and energy, and thus use cheap amino acids to minimize proteome costs. |

| Synthesis, not toxicity, explains evolution rates of overexpressed proteins.(Plata *et al.* 2010) | It is widely assumed that misfolded proteins are toxic by a specific mode of action. Plata et al. showed that turnover costs are more important for *E. coli* cell fate than toxicity at least for the studied proteins. |
|---|---|
| Sickle-cell disease patients display doubling of protein turnover and 20% increase in resting metabolism.(Badaloo *et al.* 1989) | Mutations in hemoglobin lead to dysfunctional, instable proteins that are compensated by enhanced turnover and synthesis. The numbers suggest that 20% of the normal human metabolic rate is spent on protein turnover, fully consistent with consensus in the field.(Hawkins 1991; Waterlow 1995) |

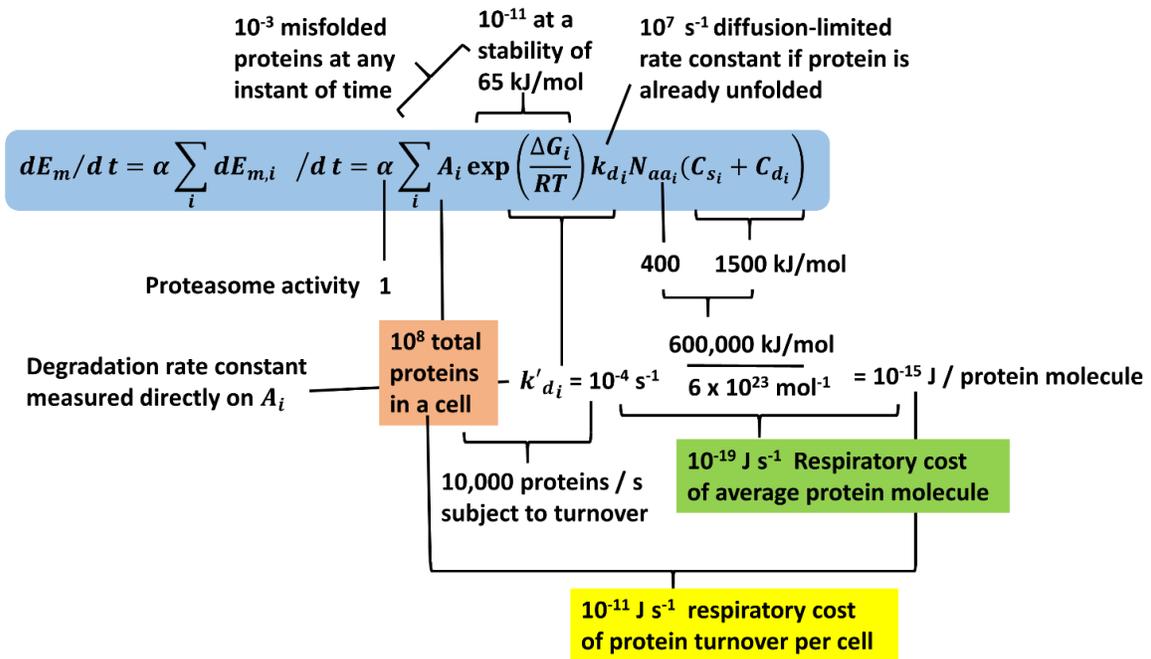

**Figure 1.** Schematic overview of order-of-magnitude terms of the PCM model. Typical values for yeast used as example. All values are subject to the well-known variations in copy numbers of individual proteins, degradation constants, length of proteins, and total number of proteins copies in a cell.

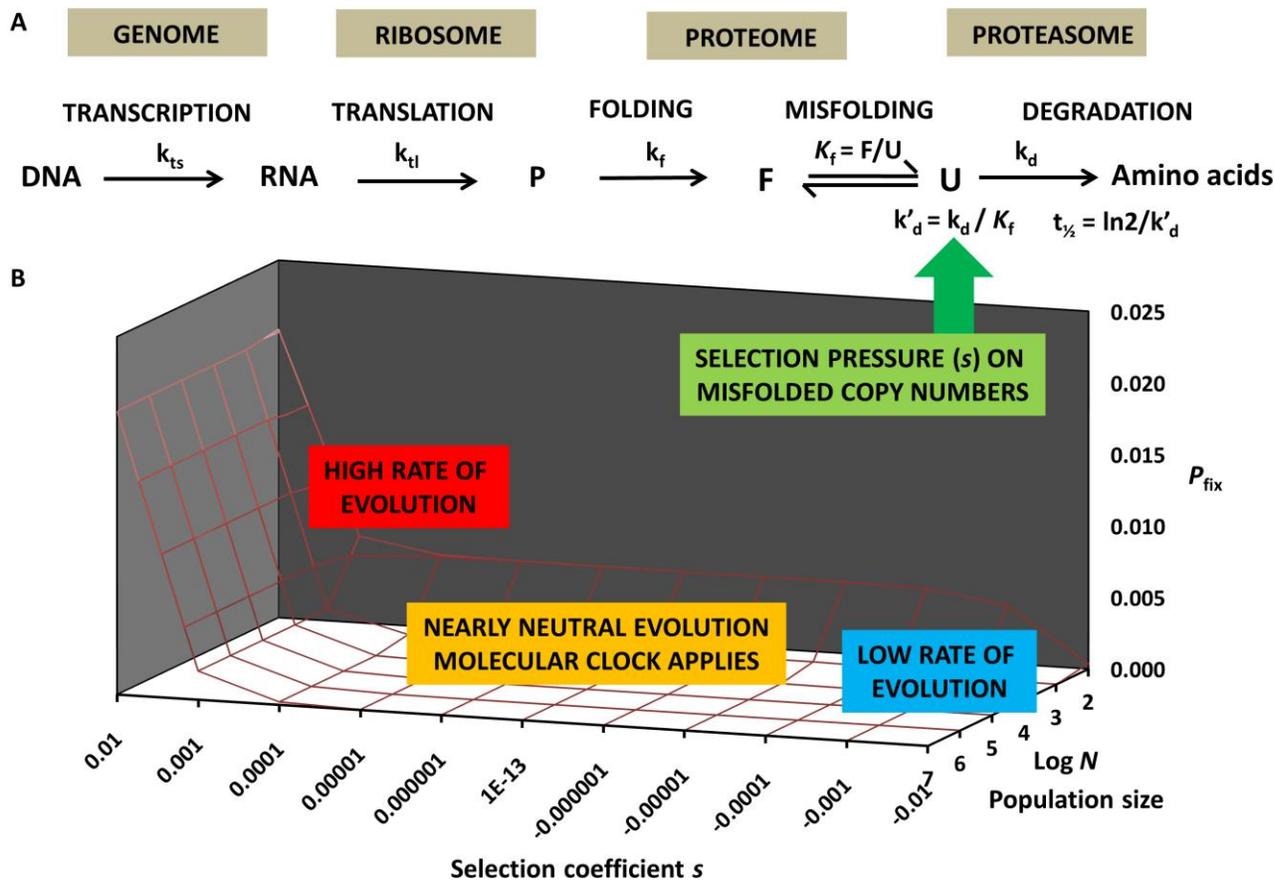

**Figure 2.** (**A**) Schematic overview of the processes of protein turnover, with the central dogma to the left and the proteome maintenance, the concern of the present paper, to the right. (**B**) Probability of fixation ($P_{fix}$) plotted against selection coefficent $s$ and log $N$ (effective population size). Beneficial mutations with $s > 0.001$ have relevant $P_{fix}$ of more than 1% for most populations. Only in very small populations ($< 100$) do other mutations get fixated ($P_{fix}$ ~1%), and neutral and slightly deleterious mutations become fixated to a similar extent until $s$ approaches $-1/N$, whence $P_{fix}$ rapidly decreases.